# Cross sections of X-ray production induced by C and Si ions with energies up to 1 MeV/u on Ti, Fe, Zn, Nb, Ru and Ta.

José Emilio Prieto[1,2], Alessandro Zucchiatti[1], Patricia Galán[1] and Pilar Prieto[3]

[1]*Centro de Micro Análisis de Materiales (CMAM), Universidad Autónoma de Madrid, 28049 Madrid, Spain*

[2]*Dpto. de Física de la Materia Condensada, IFIMAC and Instituto ¨Nicolás Cabrera¨, Universidad Autónoma de Madrid, 28049 Madrid, Spain*

[3]*Dpto. de Física Aplicada, Universidad Autónoma de Madrid, 28049 Madrid, Spain*

ABSTRACT

X-ray production differential cross sections induced by C and Si ions with energies from 1 MeV/u down to 0.25 MeV/u, produced by the CMAM 5 MV tandem accelerator, have been measured for thin targets of Ti, Fe, Zn, Nb, Ru and Ta in a direct way. X-rays have been detected by a fully characterized silicon drift diode and beam currents have been measured by a system of two Faraday cups. Measured cross sections agree in general with previously published results. The ECPSSR theory with the united atoms correction gives absolute values close to the experimental ones for all the studied elements excited by C ions and for Ta, Nb and Ru excited by Si ions. For Ti, Fe and Zn excited by Si, the basic ECPSSR theory gives better agreement, although on absolute values the gap for Ti is still large.

1-INTRODUCTION

The development of Secondary Ion Mass Spectrometry (SIMS) with incident ions at MeV energies promises to offer the community a new analytical methodology and is the object of a coordinated research effort, promoted by the International Atomic Energy Agency (IAEA) [1]. Due to the multi-technique character of Ion Beam Analysis (IBA), the development opens also the way to the combination of MeV-SIMS with other IBA techniques, in particular particle-induced X-ray emission (PIXE). Several data sets exist already for ionization and production cross-sections. However, one has to observe that lesser direct results are found for X-ray production [2-20] and that the actually available cross sections are far from being ideal, in terms of their analytical applicability to Heavy-Ion PIXE (HI-PIXE) quantification, due to the largely incomplete coverage of ions, targets and energies, something that is fully understandable, given the plethora of possible combinations. Focusing on the applicability of HI-PIXE in connection with MeV SIMS, the coordinated research has identified specific needs for cross section data and has elaborated concerted measurement protocols and cross check procedures in different laboratories. The aim is to fill to some extent the existing gaps and produce data sets convenient both as a reliable operative tool and as an experimental basis on which to test theoretical models. In this framework, we have investigated the production of X-rays induced by C and Si ions from 1 MeV/u down to 0.25 MeV/u on the following thin targets: $TiO_2$, $Fe_3O_4$, $ZnO$, $Nb_2O_5$, $RuO_2$ and $Ta_2O_5$ grown on different substrates.

2-EXPERIMENTAL SET-UP

Cross sections have been measured using the 5 MV tandem accelerator at the Centro de Micro-Análisis de Materiales (CMAM) of the Universidad Autónoma de Madrid, Spain. We have used a set of targets to measure HI-PIXE cross sections, to determine the X-ray detector energy calibration and efficiency as well as the reduction of charge on sample due to a transmission FC, as explained below. Target thicknesses have been extracted from the analysis of elastically backscattered 2 MeV alpha particles, performed with the help of the SIM-NRA program. Results are reported in Table I. The estimated error of the target thicknesses is around 2.5%.

We have operated in an experimental station equipped with two Faraday cups (FC): an insertion FC and a transmission FC, both with a secondary electron suppressor. The transmission FC consists of a fast spinning C-shaped metal wire, to brush an area that contains the beam, in electric contact with a surrounding cylinder to collect the secondary electron eventually emitted by the wire or repelled by the suppression ring. This allowed the use of thin targets on thick substrates while measuring the collected charge in a reliable way. The insertion FC was used to determine the net charge arriving on the target immediately before and after the irradiation, while the transmission FC was used to record possible current fluctuations which were eventually corrected. The transmission FC intercepts 7.1 percent (standard deviation 0.

2 percent) of the beam going to sample, as determined using a 2 MeV proton beam. A plot of the beam currents during a measurement is given in figure 1. The error of the measured charge depends on the stability of current and its correction: its average is ± 2% and the maximum ± 7%.

Proton PIXE on the targets of Table I (plus a CsBr Micromatter standard) has been performed to determine the detector energy calibration, the stability of which was checked by a $^{55}$Fe source before the ion beam measurements. At the same time we have determined the absolute efficiency of the X-ray detector using the proton-induced X-ray production cross sections given by the GUPIX program, which are tabulated from validated sets of data. X-rays were detected by a KETEK AXAS-A 10mm$^2$ silicon drift detector (SDD), placed at 120 degrees from the beam direction and protected by a convenient absorber (500 microns polyethylene terephthalate PET) from the impact of backscattered ions and protons. At low energy (see figure 2), the efficiency is reduced because of the thick proton and ion absorber. At high energy, it drops again because of photons crossing the detector's active thickness (SDD 500 microns) without interaction. The curve has been fitted by a cubic function. The error of the efficiency is on average 4% (min 2.8- max 7.2); the difference with the fitting curve is on average 7%. Point values have been used in our calculations.

Currents between 5 and 65 nA have been used for C ions in charge states ranging from 2 to 4 and between 20 and 60 nA for Si ions in charge states 3 to 5. The dead time could be maintained generally below 1%. Preamplifier signals have been processed by an ORTEC 572A amplifier and a Fast Comtec 7072 dual analog-to-digital converter. Spectra have been collected and analyzed with a Fast Comtec MP3 multi-parameter system. The statistical error for Ti, Fe, Zn K$_\alpha$ peaks and Ta L$_\alpha$ peak is always well below 1%. For Nb and Ru, as well as for Ti and Zn, it can reach, at the lowest measured energies, a value of 9%.

3-DIFFERENTIAL CROSS SECTIONS

The differential cross sections for the reaction channels mentioned above are given by:

$$\frac{d\sigma_X(E_I,\theta)}{d\Omega} = \frac{Y_X(E_I,\theta)}{N_I N_T \varepsilon_{abs}(E_X) \cdot 4\pi} \cdot f_{corr} \qquad N_I = \frac{Q}{ne} \qquad N_T = \frac{N_0 \rho dx}{A} \qquad (1)$$

where $E_I$ is the incoming ion energy, $N_I$ is the number of incident ions, $Q$ is the total collected charge, $n$ is the ion charge state, $N_T$ is the number of target atoms, $\rho \cdot dx$ is the target areal density, $N_0$ is the Avogadro number, $e$ is the unit charge and $Y_X(E_I, \theta)$ is the X-ray yield. All quantities on the right side have been measured and the cross section has been directly extracted from equation (1).

The correction factor $f_{corr}$ due to ion energy loss and X-ray absorption (negligible) in the thin targets is computed from:

$$f_{corr} = 1 - \frac{\sigma(E_{in})\rho dx}{\int_{E_{in}}^{E_{out}} \sigma(E) \cdot S^{-1}(E) dE} \qquad (2)$$

using for $\sigma(E)$ a fit of the ECPSSR values **[21-23]** and for $S(E)$ a fit of the SRIM energy loss curve **[24]**. As a matter of fact the correction is only appreciable and has been applied therefore only at the lowest energies: for C ions at 4 and 6 MeV and for Si ions at 8, 12 and 16 MeV. The applied corrections are shown in Table II.

The actual experimental-set-up allowed us to measure both the $K_\alpha$ and $K_\beta$ yields in Ti, Fe, Zn at all energies; they were added to give the K cumulative differential cross-section. Since the lower cross-section makes the background subtraction on the $K_\beta$ line more critical, we could measure for Nb and Ru only the $K_\alpha$ intensity and only at the highest energies. The $L_\alpha$ line of Ta has been integrated at all energies. The same procedure for the other L lines would imply a complete fitting of the corresponding energy regions which requires further attention due to the peak energy shifts and broadening. For our cross sections we estimate a global statistical error of about ±5% for Ti, Fe, Zn, Ta and of about ±10% for Nb and Ru.

The results for the Ti, Fe and Zn cumulative differential K cross-sections are given in figures 3 and 4 for C and Si ions respectively. Recent data for Ti by Msimanga and coworkers **[25]** show a similar trend as ours with a difference of a few tens percent except al the lowest common energy where they are even bigger. The results for Ta, Nb and Ru differential cross-sections are given in figures 5 and 6 for C and Si ions respectively.

As shown in figure 7, the basic ECPSSR **[21-22]** theory (ISICSoo code **[23]**) gives much higher values than the united atom (UA) correction. This holds also for Si ions. C-induced cross sections are described reasonably well by the ECPSSR-UA theory for Ti, Fe and Zn (figure 3). For Ta, Nb and Ru the agreement shown by figure 5 is equally satisfactory; however it has to be remarked that the experimental data correspond only to partial $L_\alpha$ and $K_\alpha$ cross sections. In particular for Ta we expect a significant increase of the experimental curve (see comments above). For the case of Si we observe (figure 4) that it is the basic ECPSSR theory that gives the closest agreement with the Ti, Fe and Zn data, although the difference is quite high, both for Ti and Zn. The Msimanga data for Ti **[25]** show again a similar trend as ours with a maximum difference of only 7%. For Ta, Nb and Ru both the basic and the united atoms ECPSSR theories give a reasonable agreement with the experiment, being the ECPSSR values higher than the data and the ECPSSR-UA lower. Only the latter are shown in figure 6. Comparing our data with previous ones on Fe **[7,13]** in the case of C (figure 7), we observe consistency with those of Fazinic et al. **[7]**, assuming similar statistical errors, but large discrepancies with those of de Lucio et al. **[13]**, increasing with energy. Furthermore, Fazinic data and ours closely follow the trend marked by the theory.

## 4-CONCLUSIONS

PIXE cross sections have been determined for the cases of excitation with C and Si ions with energies of 1 MeV/u and below, on a set of relatively thin targets from measured quantities: the collected charge, the reaction yield, the target thickness, the detector absolute efficiency. Correction factors due to the energy loss in the targets have been applied. A transmission Faraday cup has been used to check and eventually correct for current fluctuations. The obtained experimental values are in general comparable to previously determined ones being the best case that of C on Fe. The ECPSSR theory with the united atoms correction reproduces the trend of our data and gives absolute values in reasonable agreement with the experimental ones for all the elements excited by carbon ions and for Ta, Nb and Ru excited by silicon ions. For Ti, Fe and Zn excited by silicon, the basic ECPSSR theory seems more appropriate, although on absolute values the gap theory-experiment for Ti is still large. We may observe that the measurement protocol must be improved for Nb, Ru and Ta to include other X-ray lines in the analysis. Measurements for other ions (O, Cl, Cu, Br) are in progress following the recommendation issued by the coordinated research project.

## ACKNOWLEDGMENTS

We gratefully acknowledge the support of the CMAM technical staff. This work was performed within the IAEA coordinated research project (CRP) #F11019: "Development of molecular concentration mapping techniques using MeV focused ion beams" and partly supported by Project No. MAT2014-52477-C5-5-P of the Spanish MINECO. We are deeply indebted to the colleagues of the CRP for the fruitful informative discussions and for supplying targets.

TABLES

| Target | Analyte Thickness [μg/cm2] | Target Thickness [μg/cm2] | Density [g/cm3] |
|---|---|---|---|
| TiN/Si | 11.19 | 14.52 | 4.82 |
| TiN-C/Si | 19.09 | 27.97 | 4.82 |
| $TiO_2$/Si | 11.55 | 19.38 | 4.23 |
| ZnO/Si | 22.93 | 29 | 5.61 |
| $Fe_3O_4$-C/Si | 64.02 | 85.76 | 5.17 |
| $RuO_2$/Sigradure | 10.98 | 15.57 | 6.97 |
| $Ta_2O_5$/Sigradure | 15.92 | 19.78 | 8.18 |
| $Nb_2O_5$/sigradure | 14.31 | 21.05 | 4.6 |

**Table I.** Thickness of the used targets as determined by 2 MeV alpha backscattering. Targets TiN_C and $Fe_3O_4$_C were produced at CMAM.

| | $f_{corr}$ [%] | | | | |
|---|---|---|---|---|---|
| Target | C 4 MeV | C 6MeV | Si 8 MeV | Si 12 MeV | Si 16 MeV |
| $TiO_2$ | 18 | 6 | 20 | | |
| TiN | 22 | 5 | 16 | | |
| TiN-C | 23 | 7 | 18 | 2 | |
| $Fe_2O_3C$ | 33 | 10 | 31 | 13 | 8 |
| ZnO | 19 | 8 | 16 | 5 | 8 |
| $Nb_2O_5$ | 15 | 1 | 11 | | |
| $RuO_2$ | 2 | 3 | 6 | 3 | |
| $Ta_2O_5$ | 13 | | 9 | | |

**Table II.** Correction factors calculated from equation (2) for the targets and in the cases that it was applied to.

FIGURES

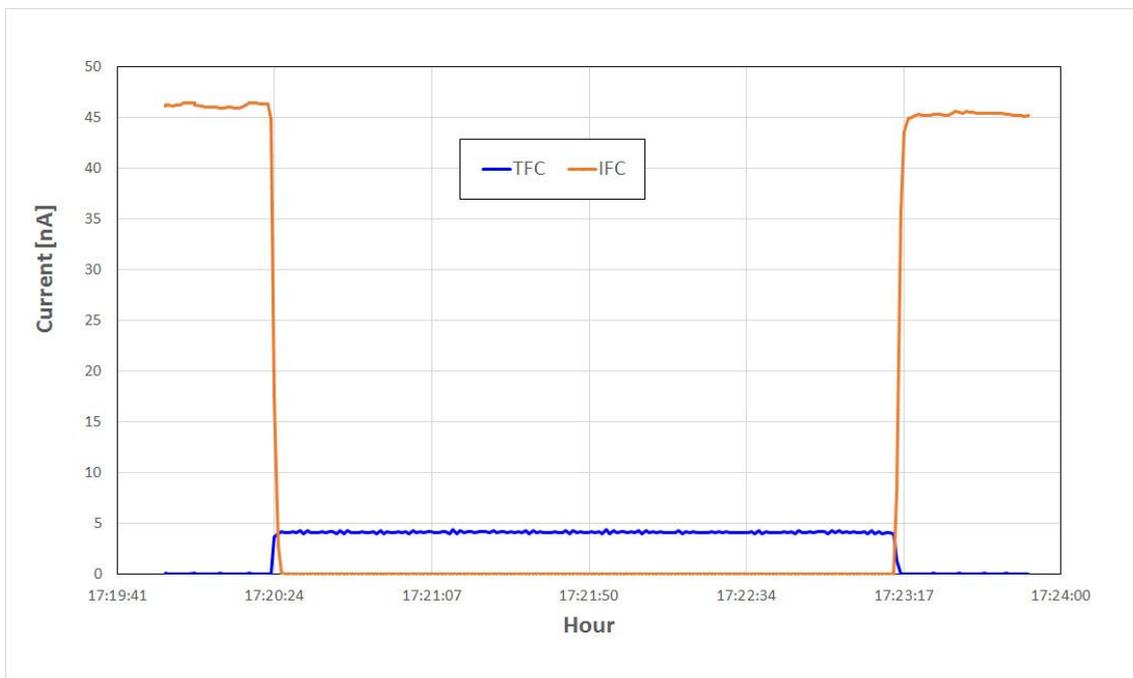

**Figure 1.** The currents recorded by insertion and transmission Faraday cups during a typical measurement.

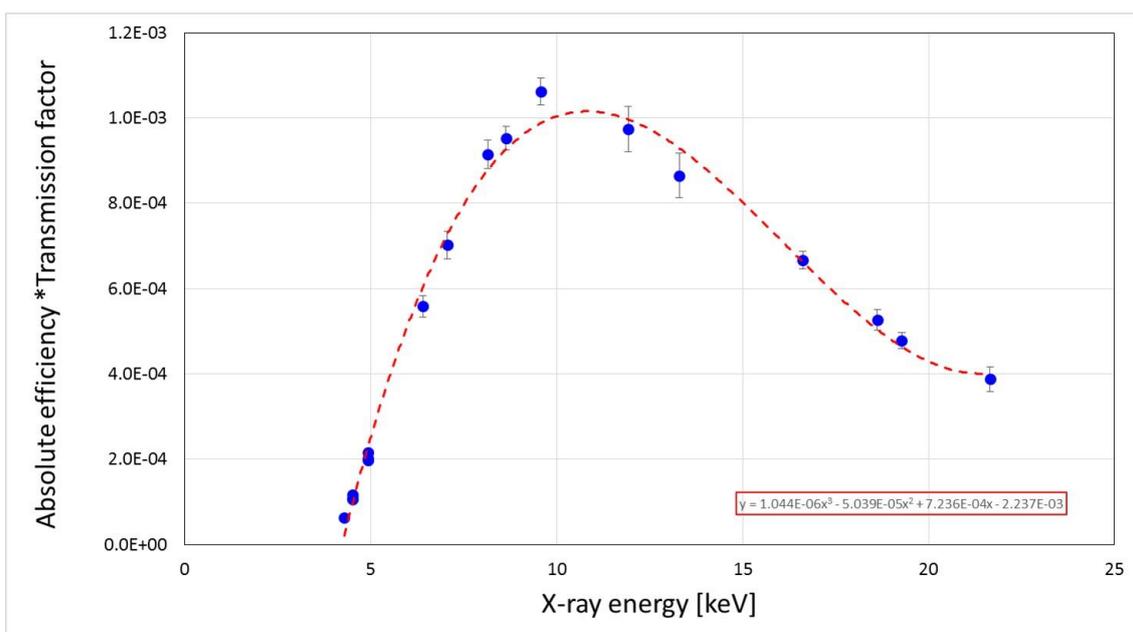

**Figure 2.** The absolute efficiency of the detector, multiplied by the filter transmission, was determined by PIXE with 2 MeV protons, as described in the text.

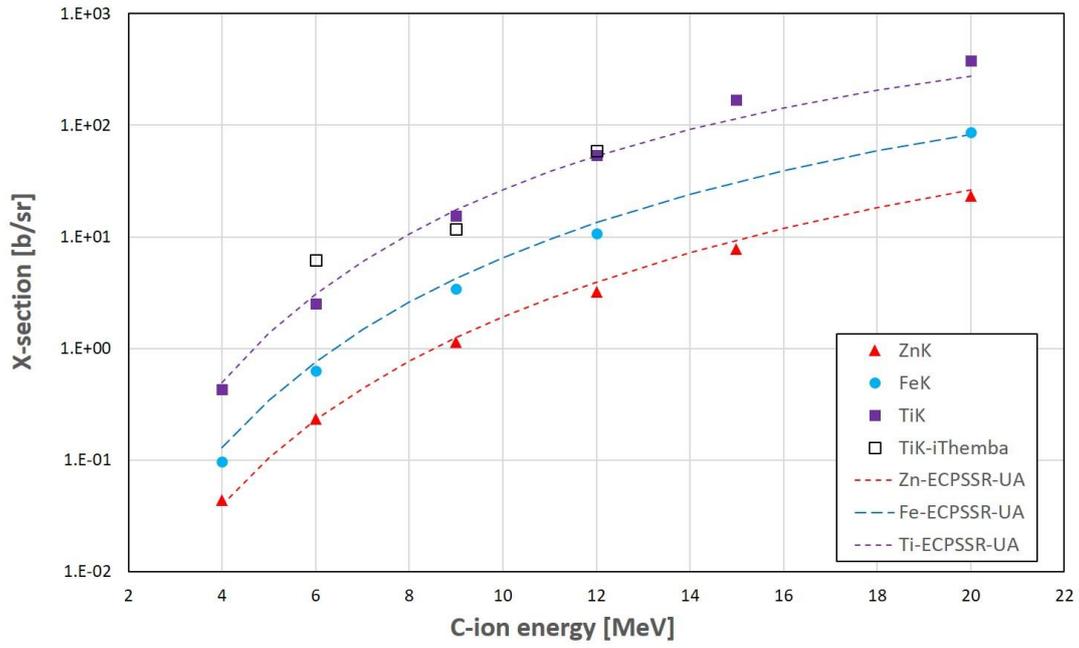

**Figure 3.** Cumulative K differential cross section obtained from C ions on Ti, Fe and Zn.

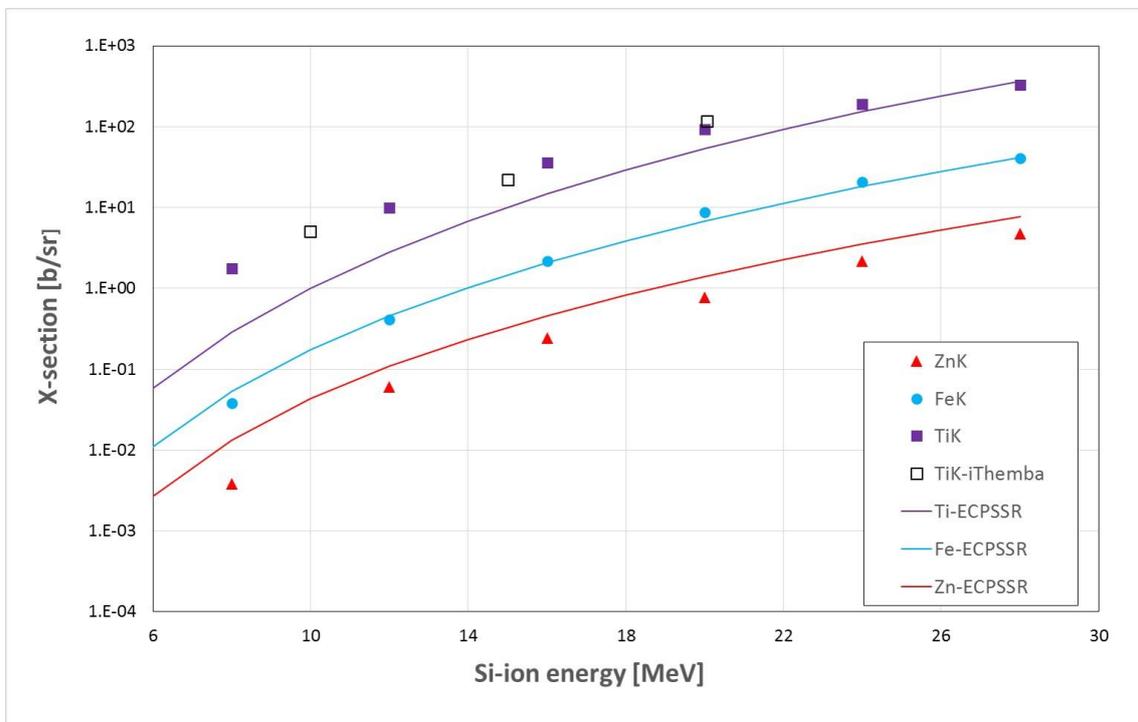

**Figure 4.** Cumulative K differential cross section obtained from Si ions on Ti, Fe and Zn.

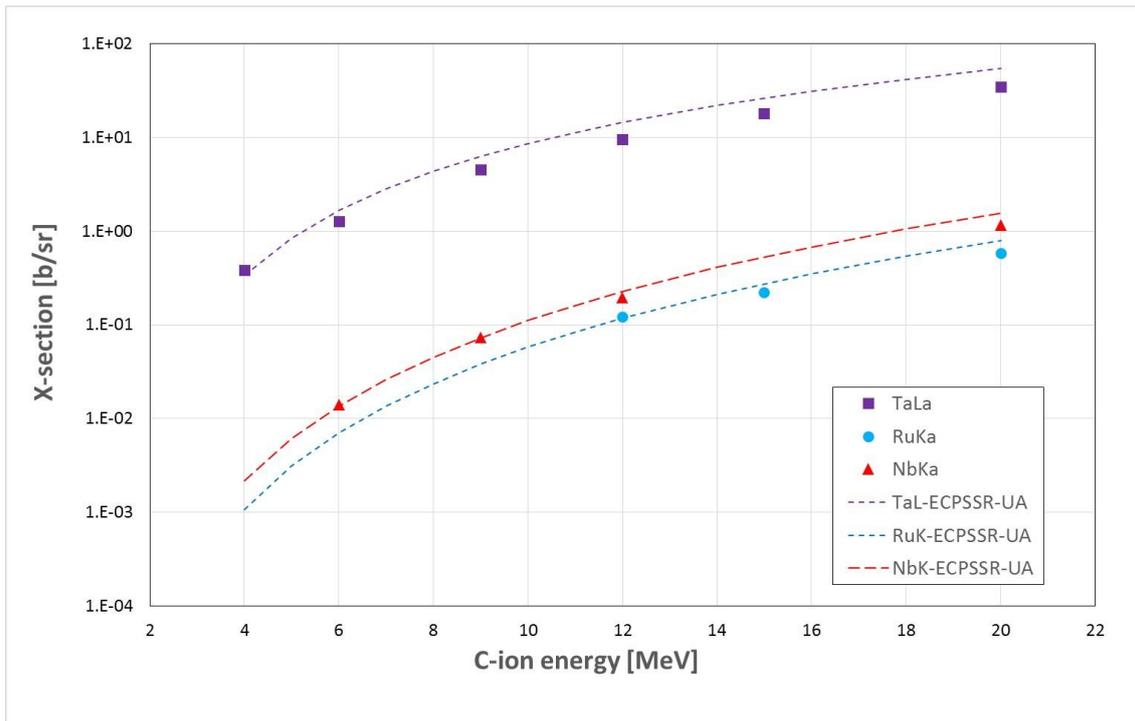

**Figure 5.** Partial differential cross section obtained from C ions on Ta (Lα line), Rb and Nb (Kα line).

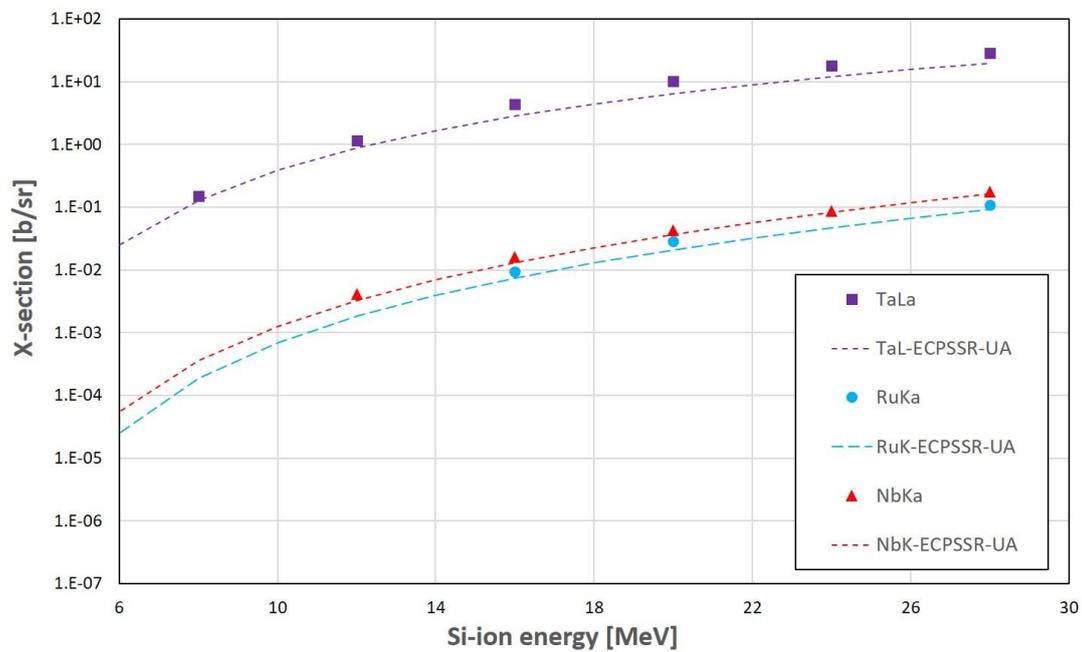

**Figure 6.** Partial differential cross section obtained from Si ions on Ta (Lα line), Rb and Nb (Kα line).

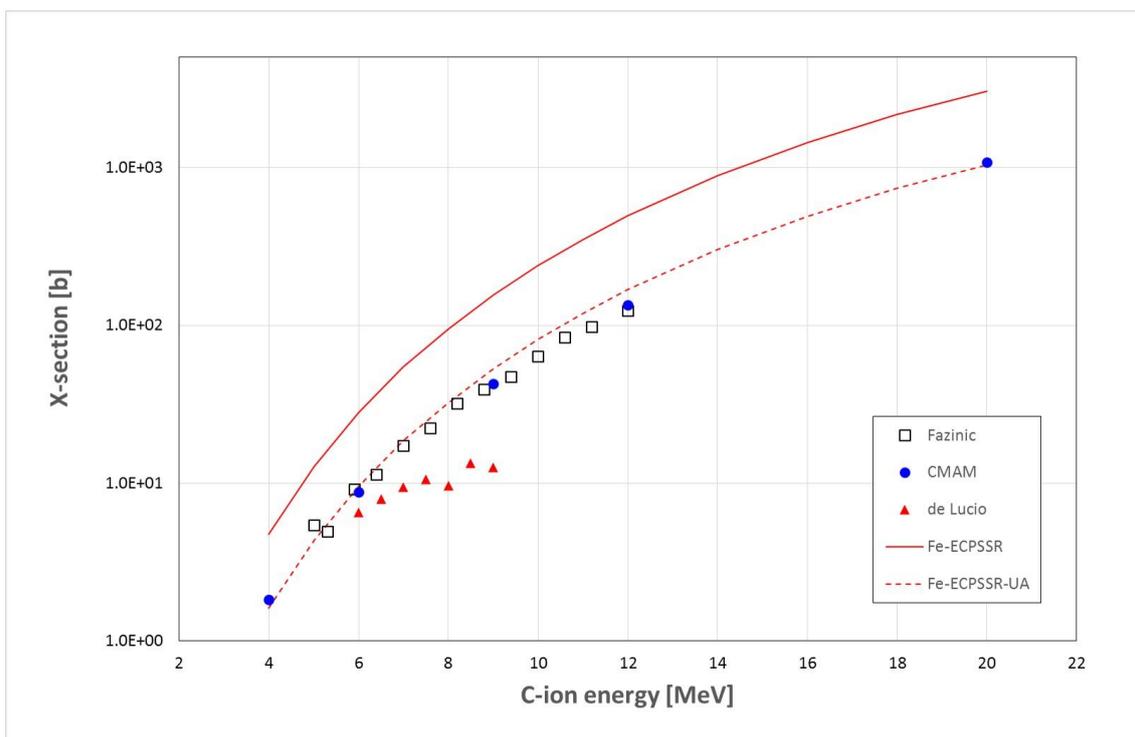

**Figure 7.** Total cross section for K X-ray production induced by C ions on Fe, compared to the results of references [7] and [13] and with the ECPSSR and ECPSSR-UA theories [21].